\documentstyle[prl,epsf,aps]{revtex}
\begin{document}
\newcommand{\ve}[1]{\mbox{\boldmath $#1$}}
\newcommand{\muk}{\mu{\text{K}}}
\newcommand{\mum}{$\mu_\text{m}$ \ }
\newcommand{\mfp}{l_{\text{mfp}}}
\newcommand{\dnu}{\Delta\nu_{\text{rf}}}

\draft
\twocolumn[\hsize\textwidth\columnwidth\hsize\csname @twocolumnfalse\endcsname

\title {Excitations of a Bose-Einstein Condensate at Non-Zero
Temperature: A Study of Zeroth, First, and Second Sound}

\author{D.M. Stamper-Kurn, H.-J. Miesner, S. Inouye, M.R. Andrews, 
and W. Ketterle}
\address{Department of Physics and Research Laboratory of 
Electronics, \\
Massachusetts Institute of Technology, Cambridge, MA 02139}
\date{Submitted to Phys. Rev. Lett. January 25, 1998} \maketitle

\begin{abstract}
Collective excitations of a dilute Bose
gas were probed above and below the Bose-Einstein condensation
temperature.
The temperature dependencies of the frequency and damping
rates of condensate
oscillations indicate
significant interactions between the condensate and the thermal cloud.
Hydrodynamic oscillations of the thermal cloud were observed,
constituting first sound.
An antisymmetric dipolar
oscillation of the thermal cloud and the condensate was studied,
representing the bulk flow of a superfluid through the normal fluid.
The excitations were observed {\it in situ} using non-destructive
imaging techniques.
\end{abstract}
]
\vskip1pc
Bose-Einstein condensation in dilute atomic gases
~\cite{ande:95,davi:95,brad:97} has provided a testing ground
for well-developed many-body theories of quantum fluids.
Many studies have focused
on the collective
excitations of such gases, which fully describe their dynamics
and transport properties such as superfluidity.
Experiments have studied
low-lying collective excitations over a range of
temperatures~\cite{jin:96,mewe:96b,jin:97} and
higher-lying modes~\cite{andr:97b}.
Zero-temperature findings have agreed well with theoretical
predictions
based on a mean-field description of the weak interatomic
interactions~\cite{edwa:96,stri:96,sound_theory}.
However, the behavior at non-zero temperature, which
involves interactions between the condensate and the thermal cloud,
is not fully understood,
pointing to the need for new theoretical developments.

The physical nature of collective excitations depends on the 
hierarchy
of three length scales:
the wavelength of the excitation $\lambda$,
the healing length $\xi$ which is determined by the condensate
density,
and the mean-free path $\mfp$ for collisions between the collective 
excitation
and other excitations which comprise the thermal cloud.
The collisionless regime, defined by
$\lambda \ll \mfp\ $, occurs at zero temperature and for low densities of the thermal
cloud.
There,
free-particle excitations are obtained at short wavelengths ($\lambda 
\ll \xi$)
while phonon-like excitations known as zeroth sound are obtained at
long wavelengths ($\lambda \gg \xi$).
For excitations where $\lambda$ is comparable to the size of the sample, this
latter condition defines the Thomas-Fermi regime.
Experiments have been performed either in the limit of 
zeroth sound
excitations~\cite{mewe:96b,andr:97b}, or intermediate to the two 
limits~\cite{jin:96,jin:97}.

At higher thermal densities, when
$\lambda \gg \mfp\ $
collective excitations become hydrodynamic in nature,
and one expects two phonon-like 
excitations
known as first and second sound.
The presence of two hydrodynamic modes is similar to the case
of superfluid $^4$He.
However, while in superfluid $^4$He second sound is an entropy wave 
with no density oscillation,
in a dilute Bose gas both first 
and second sound describe density waves.
First sound, which is defined as having the larger speed of sound,
involves mainly the thermal cloud, whereas
second sound is similar in character to zeroth sound and is confined
mainly to the condensate~\cite{first_second,zare:97}.

In this Letter, we study collective excitations at non-zero temperature,
and extend earlier work~\cite{jin:97} in several ways.
First, we study zeroth sound condensate oscillations entirely
in the Thomas-Fermi regime.
In contrast, results of the earlier study were complicated by the transition
of the condensate from the Thomas-Fermi to the free-particle
regime with increasing temperature.
Second, near the critical temperature we approached the hydrodynamic limit
and observed the onset of hydrodynamic excitations analogous to first sound.
Finally, we observed a new out-of-phase dipolar oscillation
of the condensate and the thermal cloud, analogous to the out-of-phase
second sound mode in liquid helium~\cite{zare:97}.

The excitations were probed generally in three steps.
First, as described in previous work~\cite{mewe:96a}, we produced a
magnetically confined,
ultracold gas of atomic sodium in a well-determined equilibrium state.
This state was controlled by adjusting the final frequency 
$\nu_{\text{rf}}$ used
in the rf evaporation by which the clouds were cooled.
The clouds were cigar-shaped with weak confinement along one direction (axial),
and tight confinement in the other two (radial).
Second, the cloud was manipulated with either time-dependent magnetic fields 
or
off-resonant
light to excite low-lying collective modes.
Finally, the cloud was allowed to oscillate freely and probed 
{\it in situ}
with repeated, non-destructive phase-contrast imaging~\cite{andr:97b}.

To accurately characterize the magnetic trapping potential, we excited 
center-of-mass
oscillations
of the cloud in the axial direction by sinusoidally varying
the position of the trap center.
By observing the subsequent motion of the cloud,
we measured the axial trapping frequency as $\nu_z = 16.93(2)$~Hz,
while the radial frequency was estimated to be $230$ Hz
based on earlier measurements.

The $m = 0$ quadrupolar modes of the condensate and the thermal 
cloud~\cite{mewe:96b}
were excited by a 5-cycle pulsed modulation of the axial magnetic field curvature.
Condensate oscillations were produced with driving frequencies of 25 
-- 27 Hz,
while the thermal cloud was excited at about 30 Hz.
As shown in Figure 1, the most prominent feature of the condensate 
oscillations was the change in the axial length.
Data were evaluated for oscillations with a relative amplitude of 
about 10\%~\cite{anharmonicity_footnote}.

The oscillations of each cloud were probed with 22 non-destructive 
images: one
before the excitation to characterize the initial conditions, and 
three groups of 7 images during the oscillation.
The three groups were separated by a delay time
which was varied between 1 and 200 ms according to the damping time
of the oscillation.
This method of probing gave highly accurate single-shot measurements
of oscillation frequencies and damping rates,
allowed data to be collected efficiently,
and overcame the additional fluctuations introduced when
combining data from observations
on several clouds.

Phase-contrast images were analyzed by fitting
the observed 
column densities $\tilde{n}(r,z)$~\cite{calibration_footnote}
with the function
\begin{eqnarray}
\tilde{n}(r,z) & = & h_c \, \text{max}\biggr(0, 1 - \frac{r^2}{(l_r/2)^2} -
\frac{z^2}{(l_z/2)^2}\biggr)^{3/2} + \nonumber \\
               &   & h_t \, g_2\biggl( \exp(
-\frac{r^2}{2 (\sigma_r/2)^2} -
\frac{z^2}{2 (\sigma_z/2)^2})\biggr).
\label{column_density_equation}
\end{eqnarray}
Here, $r$ and $z$ are
the radial and axial coordinates, respectively,
while $l_r$ and $l_z$ are the lengths of the condensate
and $\sigma_r$ and $\sigma_z$ are the rms diameters of the thermal 
cloud.
The function $g_2$ is defined by $g_2(x) = \sum_{i=1}^{\infty} 
x^{i}/i^{2}$.
Eq.\ ~\ref{column_density_equation} is motivated by a simple model of the
mixed cloud: a mean-field dominated condensate amid a saturated
non-interacting non-condensed Bose gas.
Yet, since all quantities were allowed to vary independently,
Eq.\ ~\ref{column_density_equation} is an almost model-independent
parametrization of a bimodal distribution:
$l_z$ is determined from the cusps of the bimodal 
distribution,
and $\sigma_z$ from the thermal tails.

The initial conditions for the oscillation were characterized by the
total number of atoms $N$, the temperature $T$,
and the chemical potential
$\mu$.
$N$ was obtained by integrating the column density,
while $T$ and $\mu$ were determined from the fits by
$k_B T = \pi^2 m \nu_z^2 \sigma_z^2$ and
$\mu = \pi^2 m \nu_z^2 l_z^2$,
where $k_B$ is Boltzmann's constant.
For mixed clouds, $T$ was determined by fitting the thermal wings alone.
The initial conditions (Fig.\  2) were varied by adjusting
$\dnu\ = \nu_{\text{rf}} - \nu_{\text{bot}}$ where 
$\nu_{\text{bot}}$
is the resonant rf frequency at the bottom of the magnetic trap
which remained constant within 20 kHz.
The Bose-Einstein condensation transition was observed at $T = 1.7 \, 
\mu\text{K}$ with
about $80 \times 10^6$ atoms.
The temperature varied linearly with $\dnu\ $, with a slope of
3.5 $\mu\text{K}$/MHz,
and was measurable down to $0.5 \, \mu\text{K}$.
Reliable determinations of $l_z$ (and $\mu$) were obtained only below
$\dnu\ = 350$ kHz.
For low $\dnu\ $, $\mu / k_B \simeq 380 \, \text{nK}$, corresponding
to a condensate of $15 \times 10^6$ atoms with a maximum density of
$n_0 = 3.8 \times 10^{14} \text{cm}^{-3}$ using
$\mu = n_0 \cdot 4 \pi \hbar^2 a / m$ with the scattering 
length $a = 2.75 \, \text{nm}$.

The lengths $l_z$ and $\sigma_z$ were fit independently
to decaying sinusoidal 
functions of time.
We thus determined the frequency and damping rate
of two distinct oscillations.
We observed no evidence for coupling between the oscillations in $l_z$ and in $\sigma_z$,
allowing one to consider the excitations as nearly isolated 
oscillations
of the thermal cloud and of the condensate, respectively.

The thermal cloud oscillated at a frequency of
about 1.75 $\nu_{z}$ with a damping rate of
about 20 $\text{s}^{-1}$, both above and below
the transition temperature (Fig.\ 3).
The observed frequency $\nu$ is between the predictions of
$\nu = 2 \nu_z$ in the
collisionless limit, and of $\nu = 1.55 \, \nu_z$ in the
hydrodynamic limit~\cite{hydro_mode}.
The damping rate is predicted to vanish in both the collisionless
and hydrodynamic limits, and to reach a broad maximum
of $\sim 1.4 \, \nu$ when $\nu$ is between its two
limiting values~\cite{kavo:97}, in agreement with
our observation.
The collisional mean-free path
$\mfp\ \simeq (n_T \sigma)^{-1} = 96 \, \mu\text{m} \times (T/\mu\text{K})^{-3/2}$
using the peak density of the thermal
cloud $n_T = 2.612 \, (m k_B T / 2 \pi \hbar^2)^{3/2}$,
and a collisional cross section  $\sigma = 8 \pi a_{\text{sc}}^2$.
Around the transition temperature, we find $\sigma_z \simeq 8 \, \mfp\ $.
This comparison of length scales,
the observed frequency shift away from $2 \nu_z$, and the
high damping rate all demonstrate that the collective behavior
of the thermal cloud is strongly affected by collisions.
Thus, the thermal oscillations which we observe indicate the onset
of hydrodynamic excitations, i.e. first sound.
However, the pure hydrodynamic limit, characterized by low damping, would
only be reached for even larger clouds.

We studied the quadrupolar condensate oscillations in greater detail.
Typical oscillation data (Fig.\ 4.) demonstrate that
the oscillation  has a slightly lower frequency and
is damped more rapidly at high temperature than at low temperature.
At low-temperatures, the condensate oscillation frequency approached a
limiting value of 1.569(4) $\nu_z$, close to the zero-temperature,
high-density prediction
of 1.580 $\nu_z$~\cite{stri:96}.
The slight difference between these values
may be due to non-zero temperature effects even for small $\dnu\ $.
At higher temperatures, the frequency drops below 
the low-temperature limit.
This trend might be explained by
the simple non-zero temperature application of Bogoliubov 
theory~\cite{non_zero_theory},
in which the condensate oscillates in a combination of the
external trapping potential $U(r,z)$ and a mean-field potential exerted by the
(static) thermal cloud.
Considering a thermal density
$n_T \propto e^{-U(r,z)/k_B T}$, one can estimate that the
mean-field potential shifts the collective excitation
frequencies downward by as much as 5\%.
This shift, which scales as $a / \lambda_{\text{dB}}$ where 
$\lambda_{\text{dB}}$ is the thermal deBroglie wavelength,
is consistent in magnitude with those which we observed.

Damping rates for the condensate oscillations
varied strongly with temperature, rising from a low-temperature limit 
of about 2 
$\text{s}^{-1}$ to as much as 15 -- 20 $\text{s}^{-1}$ at high 
temperatures.
Recent treatments based on Landau 
damping~\cite{damping_papers,gior:97,fedi:97} provide qualitative 
agreement with our findings.
A quantitative prediction for these damping rates~\cite{fedi:97} 
could not be checked because our data were collected for $k_B T \leq 6 \mu$, where the
high-temperature prediction might not be applicable,
while the inability to measure $T$ for low $\dnu\ $ prevents
a comparison at low temperatures.

Can the condensate oscillations be considered to be second sound?
A comparison of length scales --- $l_z \simeq 4 \, \mfp $ at high-
temperatures --- suggests that hydrodynamic effects may already be present.
However, there are no theoretical predictions regarding the transition
from zeroth to second sound with which to compare our data.
In future experiments with larger condensates, the signature of this 
cross-over may appear in the damping rate of the oscillations, which
should decrease again at high-temperatures as one reaches the 
hydrodynamic limit.

To further probe the interaction between the thermal cloud and the
condensate, we studied an excitation of a different symmetry:
the rigid-body, out-of-phase motion of the condensate and the 
thermal cloud in the
harmonic trapping potential~\cite{zare:97}.
At temperatures where the condensate fraction approaches 
50\%, both components
should participate equally in this motion.
This mode is analogous to second sound in liquid helium, where
the superfluid and the normal fluid undergo out-of-phase density 
oscillations of equal magnitude.

We excited this mode by using
off-resonant, blue-detuned
laser light, which repels atoms due to the AC Stark 
shift~\cite{davi:95,andr:97b,andr:97a}.
We produced a 3 $\muk $-high repulsive potential by focusing
40 mW of light at 514 nm to an elongated spot with $1 / 
e^2$ half-widths
of about 12 $\mu\text{m}$ and 100 $\mu\text{m}$.
After a partly condensed cloud 
was formed, the light was turned on and directed at the edge of the cloud,
where it overlapped only with the thermal cloud.
By tilting a motorized mirror, the laser beam was steered
toward and then away from the center of the 
cloud,
thereby pushing the thermal cloud in the axial direction
while not directly affecting the 
condensate.
The light was then turned off, and the cloud allowed to freely 
oscillate.

The position of the center-of-mass of the condensate (Fig.\ 5a) was 
monitored with repeated
phase-contrast images, and data from several repetitions of the 
experiment were collected
with a variable delay between excitation and probing.
The condensate motion was initially slow, and then grew to an asymptotic
sinusoidal oscillation, corresponding to the in-phase motion of the 
entire cloud (condensate plus thermal cloud) in the
magnetic trap at the trapping frequency $\nu_z$.
By subtracting the undamped center-of-mass motion
of the entire cloud, we isolated the antisymmetric oscillation
of the condensate and the 
thermal cloud (Fig.\ 5b).

The frequency of the asymmetric dipole mode of
$17.26(9)$ Hz was significantly lower than the trapping frequency
which was
$\nu_z = 18.04(1)$ Hz at that time.
This $\sim 5\%$ frequency shift is
again clear evidence of the interaction between the thermal cloud and 
the condensate.
The motion exhibited in this oscillation can be regarded as
the bulk flow of a 
superfluid through the surrounding
normal fluid.
Therefore, the damping which we observe directly measures the
dissipation of superfluid flow.
By varying the amplitude of the oscillation, one can 
study this dissipation as a function of velocity,
and obtain more detailed information about the 
superfluidity of Bose-Einstein condensed gases.
Further, the motion we observed, in which the
condensate is driven by the moving thermal cloud,
cannot be described by recent theories which
assume a stationary thermal cloud~\cite{non_zero_theory},
and requires more sophisticated
treatments~\cite{gior:97}.

In conclusion, we have studied the collective excitations 
of a dilute Bose gas at non-zero
temperatures in the Thomas-Fermi limit,
and near the hydrodynamic regime.
The hydrodynamic oscillation of the thermal cloud, corresponding to 
first sound, was indicated by measurements both
above and below the Bose-Einstein condensation transition.
The accurately determined frequency shift of the quadrupole oscillations
away from their zero-temperature limit and of the antisymmetric 
oscillation away from the trap frequency 
are measures of the forces exerted by the condensate and the thermal 
cloud on one another.
The damping of these modes sheds light on the dissipation of 
superfluid flow.

We are grateful to Allan Griffin, Jason Ho, Dan Rokhsar, and
Sandro Stringari for 
insightful discussions,
and to Dallin Durfee for
experimental assistance.
This work was supported by the Office of Naval Research, NSF, 
Joint Services Electronics Program (ARO), and the David and Lucile 
Packard Foundation.  
D.M.S.-K. acknowledges support from NSF Graduate Research Fellowships.

\begin{figure}
\epsfxsize = 4cm
\centerline{\epsfbox[0 0 154 126]{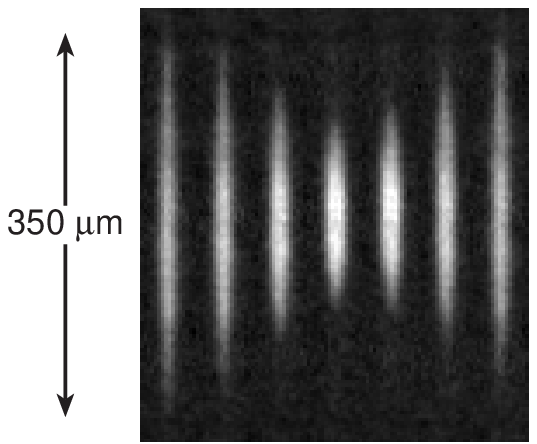}}
\begin{caption}
{{\it{In situ}} images of the $m=0$ quadrupolar condensate oscillation.
A Bose-Einstein condensate with no discernible
thermal component was imaged every 5 ms by phase-contrast imaging.
The evident change in the axial length of the condensate was
used to characterize the oscillation.
Final data were evaluated for smaller oscillation amplitudes.}
\end{caption}
\end{figure}

\begin{figure}
\epsfxsize = 6cm
\centerline{\epsfbox[0 0 351 354]{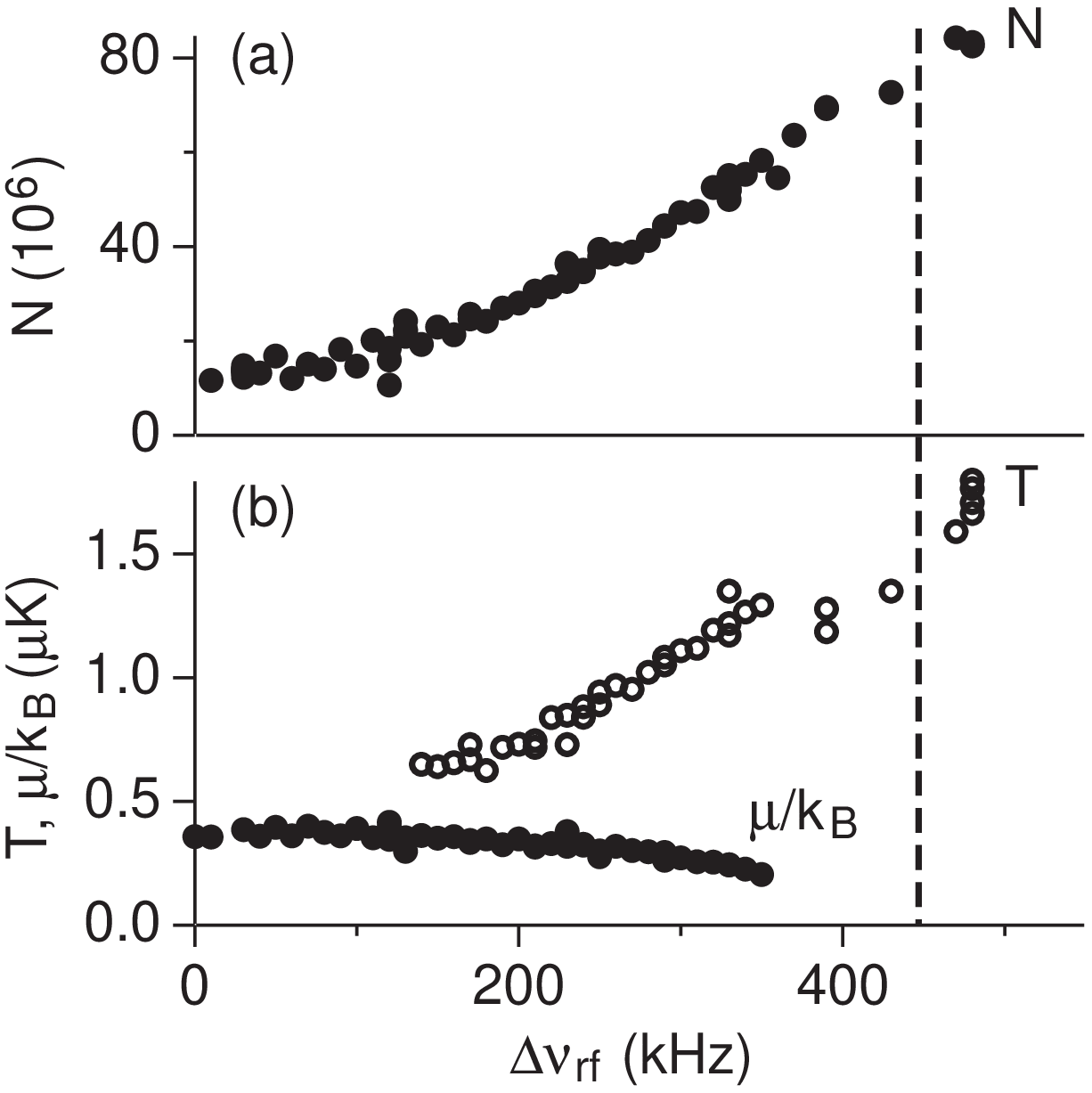}}
\begin{caption}
{Characterization of the trajectory across the Bose-Einstein 
condensation
transition.
The total number $N$ (a)
was determined by summing over the observed column densities.
The temperature $T$ (b, open circles) and chemical potential 
$\mu$
(b, closed circles) were determined from fits
to the data.
These are plotted against $\Delta \nu_{\text{rf}}$.
The dashed line indicates the observed transition temperature.}
\end{caption}
\end{figure}

\begin{figure}
\epsfxsize = 7cm
\centerline{\epsfbox[0 0 499 403]{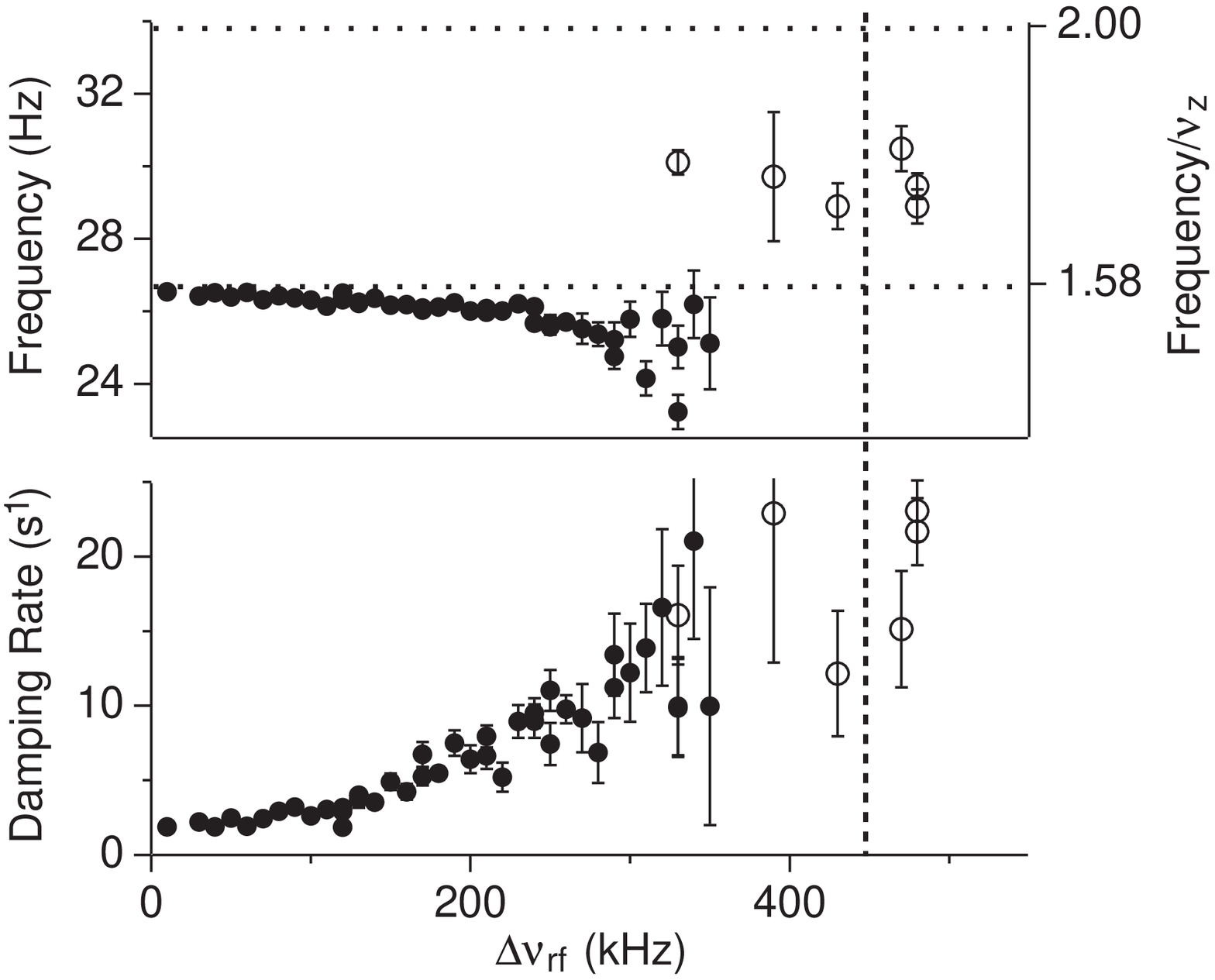}}
\begin{caption}
{Temperature dependent frequency and damping rates
of $m=0$ quadrupolar collective modes.
Points show measurements for oscillations of the
thermal cloud (open circles) and condensate (closed circles).
Horizontal dotted lines indicate the free-particle limit of 2 $\nu_z$ and the
zero-temperature condensate oscillation limit of 1.580 $\nu_z$.
The vertical dashed line marks the observed transition temperature.
}
\end{caption}
\end{figure}

\begin{figure}
\epsfxsize = 6cm
\centerline{\epsfbox[0 0 377 354]{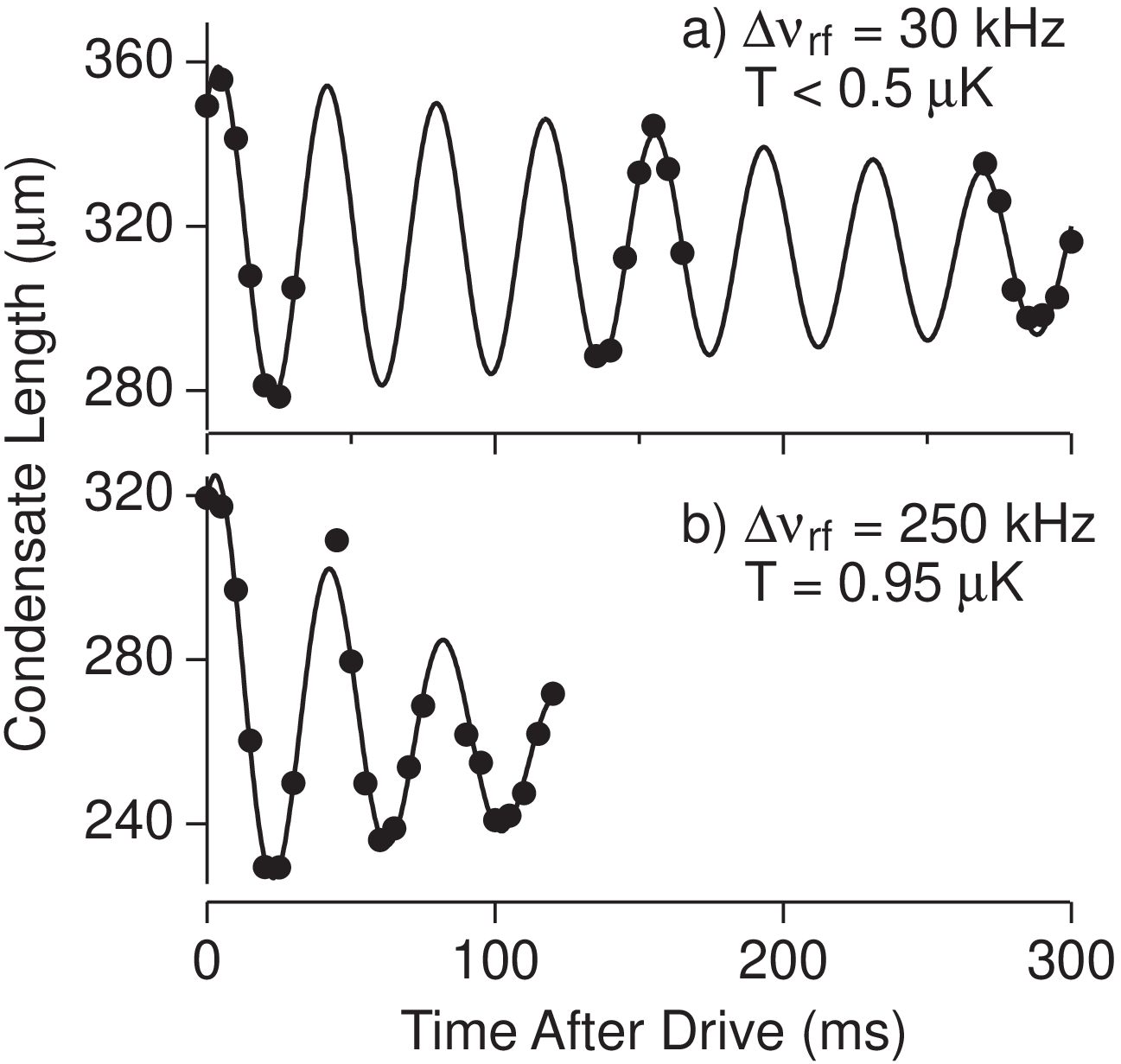}}
\begin{caption}
{Damped quadrupolar condensate oscillations at low (a) and high (b)
temperature.
Points show the axial condensate length determined from fits to
phase-contrast images.
Lines are fits to a damped
sinusoidal oscillation with a downward slope.
This slope accounted for heating during the oscillation.
The oscillation at high temperature has a slightly lower frequency, and
is damped more rapidly than at low temperature.}
\end{caption}
\end{figure}

\begin{figure}
\epsfxsize = 7cm
\centerline{\epsfbox[0 0 425 386]{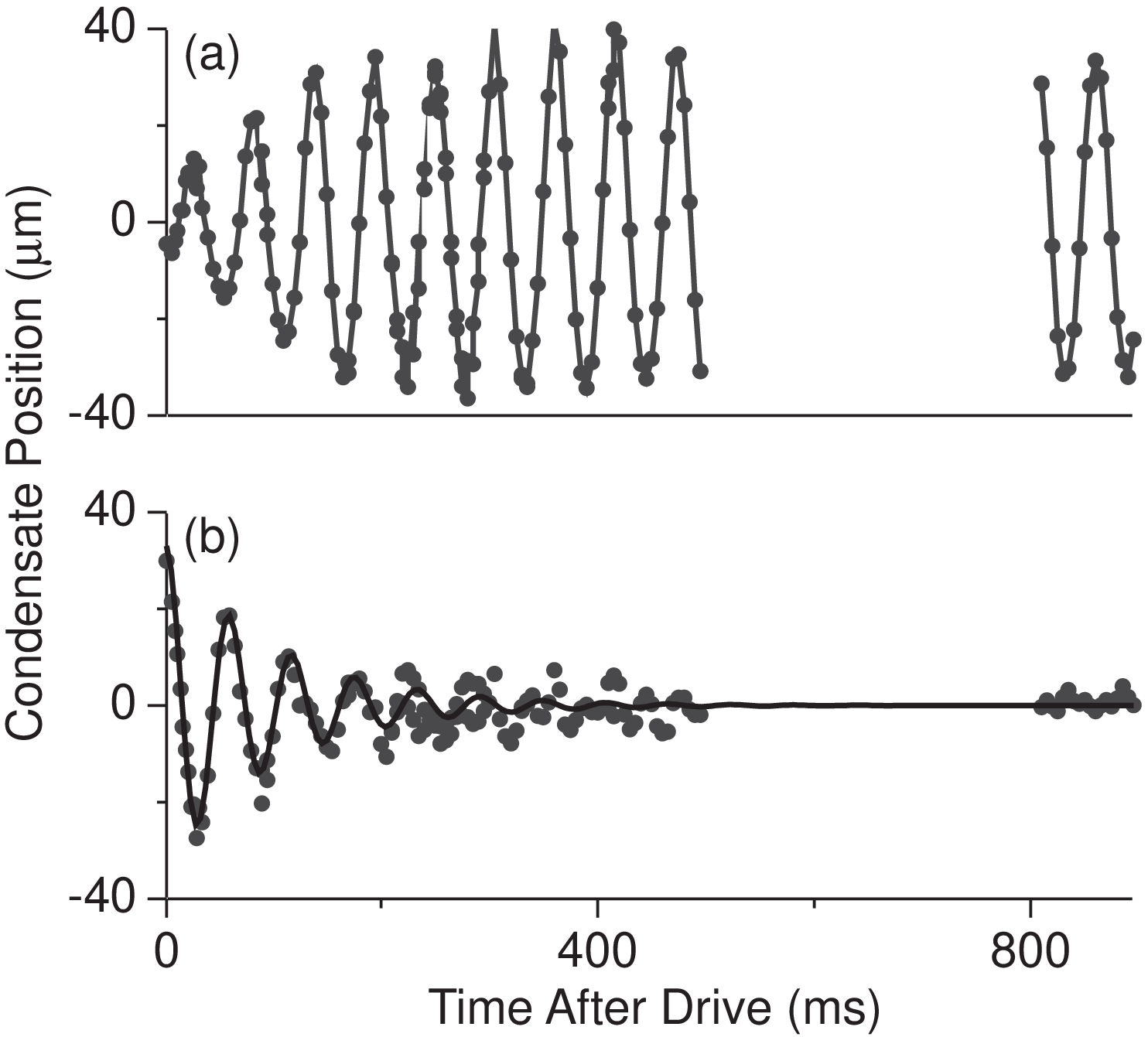}}
\begin{caption}
{Isolating the antisymmetric dipole mode.
The center-of-mass of the condensate in the trap (a) was monitored 
after the
selective displacement of the thermal cloud with a far-detuned laser 
beam.
The oscillation of the condensate relative to the center-of-mass of 
the
entire sample (b) occurred at a frequency below the trap frequency 
with a damping rate of $9(1) \text{s}^{-1}$. The initial
conditions were $T = 1 \, \muk\ $, $\mu/k_B \simeq 200 \, \text{nK}$,
and $N = 40 \times 10^6$.}
\end{caption}
\end{figure}

\begin{thebibliography}{10}
\bibitem{ande:95}
M.~H. Anderson {\it et~al.}, Science {\bf 269}, 198 (1995).

\bibitem{davi:95}
K.B. Davis {\it et~al.}, Phys. Rev. Lett. {\bf 75}, 3969 (1995).

\bibitem{brad:97}
C.C. Bradley, C.A. Sackett, and R.G. Hulet, Phys. Rev. Lett. {\bf 
78}, 985
(1997), see
also: C.C.Bradley {\it et~al.}, Phys. Rev. Lett. {\bf 75}, 1687 
(1995).

\bibitem{jin:96}
D.S. Jin {\it et~al.}, Phys. Rev. Lett. {\bf 77}, 420 (1996).

\bibitem{mewe:96b}
M.-O. Mewes {\it et~al.}, Phys. Rev. Lett. {\bf 77}, 988 (1996).
~\label{mewe:96b_label}

\bibitem{jin:97}
D.S. Jin {\it et~al.}, Phys. Rev. Lett. {\bf 78}, 764 (1997).
\label{boulder_finite_T}

\bibitem{andr:97b}
M.R. Andrews {\it et~al.}, Phys. Rev. Lett. {\bf 79}, 553 (1997).
\label{sound_paper}

\bibitem{edwa:96}
M. Edwards {\it et~al.}, Phys. Rev. Lett. {\bf 77}, 1677 (1996).

\bibitem{stri:96}
S. Stringari, Phys. Rev. Lett. {\bf 77}, 2360 (1996).

\bibitem{sound_theory}
E.~Zaremba, Phys. Rev. A {\bf 57}, 518 (1998);
G.M. Kavoulakis and C.J. Pethick, preprint cond-mat/9710130.

\bibitem{first_second}
T.D. Lee and C.N. Yang, Phys. Rev. {\bf 113}, 1406 (1959);
A. Griffin, {\it Excitations in a Bose-Condensed Liquid} (Cambridge 
University Press, Cambridge, 1993) and refs.\ therein;
A. Griffin and E. Zaremba, Phys. Rev. A {\bf 56}, 4839 (1997);
V.B. Shenoy and T.-L. Ho, preprint cond-mat/9710274.
Several authors note that the character of first 
and second sound is interchanged at very low temperatures.

\bibitem{zare:97}
E. Zaremba, A. Griffin, and E. Nikuni, preprint cond-mat/9705134.
\label{griffin_mode}

\bibitem{mewe:96a}
M.-O. Mewes {\it et~al.}, Phys. Rev. Lett. {\bf 77}, 416 (1996).

\bibitem{anharmonicity_footnote}
At a relative amplitude of 
$\approx$ 10\%, the condensate oscillation frequency
approached its low amplitude limit while the 
frequency rose
by as much as 1 Hz at an amplitude of 50\%.

\bibitem{calibration_footnote}
A phase-contrast image gives the optical phase $\phi$ accrued by
off-resonant light passing through a dense medium.
For our probe detuning, wavelength, and polarization, the
column density
$\tilde{n} = \phi \cdot 6.2 \times 10^{11} \, \text{cm}^{-2}$.

\bibitem{hydro_mode}
A. Griffin, W.-C. Wu and S. Stringari, Phys. Rev. Lett. {\bf 78}, 
1838 (1997);
Yu. Kagan, E.L. Surkov and G.V. Shlyapnikov, Phys. Rev. A {\bf 55}, 
R18 (1997).

\bibitem{kavo:97}
G.M. Kavoulakis, C.J. Pethick, and H. Smith, preprint cond-mat/9710130.

\bibitem{non_zero_theory}
D.A.W. Hutchinson, E. Zaremba, and A. Griffin, Phys. Rev. Lett. {\bf 
78},
1842 (1997);
R.J. Dodd, M. Edwards, C.W. Clark, and K. Burnett, Phys. Rev. A
{\bf 57}, R32 (1998).

\bibitem{damping_papers}
W.V. Liu, Phys. Rev. Lett. {\bf 79}, 4056 (1997);
L.P. Pitaevskii and S. Stringari, preprint cond-mat/9708104;

\bibitem{gior:97}
S. Giorgini, preprint cond-mat/9709259.

\bibitem{fedi:97}
P.O. Fedichev, G.V. Shlyapnikov, and J.T.M. Walraven,  preprint cond-mat/9710128.

\bibitem{andr:97a}
M.R. Andrews {\it et~al.}, Science {\bf 275}, 637 (1997).

\end{thebibliography}
\end{document}